\documentclass{ws-p8-50x6-01}

\newcommand{\shat}{\hat{s}^2}

\newcommand{\rwhat}{\Delta\hat{r}_W}
\newcommand{\rzhat}{\Delta\hat{r}_Z}
\newcommand{\dgama}{\hat\Delta_\gamma}

\newcommand{\etal}{{\em et al.}}

\begin{document}

\title{\uppercase{Constraining Electroweak Physics}\footnote{Talk presented at
the 2nd International Conference on String Phenomenology 2003, Durham, England,
July 29 -- August 4, 2003. The results presented here have been updated and 
differ from those shown at the conference.}}

\author{Jens Erler}

\address{Inst.\ de F{\'\i}sica, Univ.\ Nacional Aut\'onoma de M\'exico,
01000 M\'exico, D.F., M\'exico\\
E-mail: erler@fisica.unam.mx}


\maketitle

\abstracts{I summarize the status of the Standard Model after the 2003 summer 
conferences.}

\begin{figure}[t]
\epsfxsize=24pc 
\rotatebox{270.0}{\epsfbox{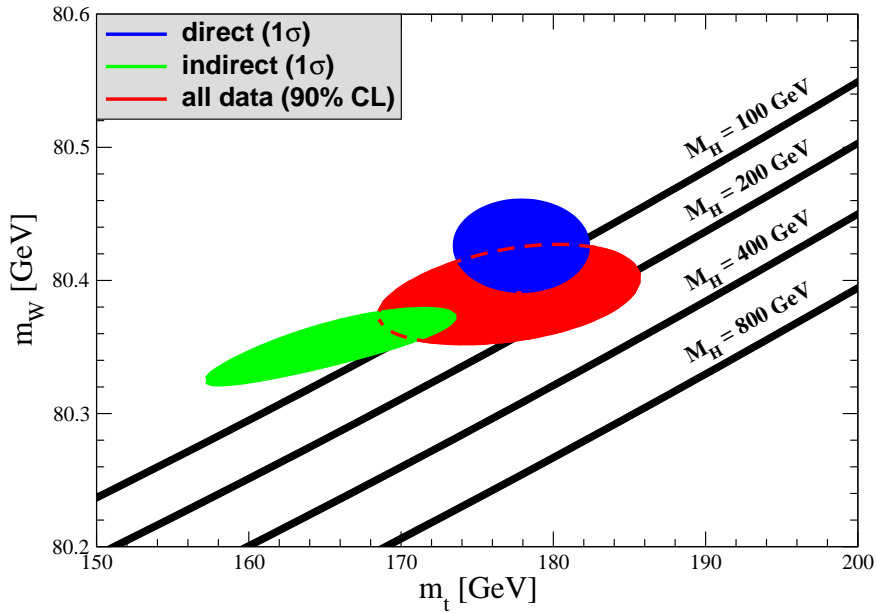}}
\caption[]{One-standard-deviation (39.35\%) region in $M_W$ as a function of 
$m_t$ for the direct and indirect data, and the 90\% CL region 
($\Delta \chi^2 = 4.605$) allowed by all data. The Standard Model (SM) 
prediction for various values of $M_H$ is also indicated. The widths of 
the $M_H$ bands reflect the theoretical uncertainty from $\alpha (M_Z)$. 
The direct $m_t = 177.9 \pm 4.4$~GeV is the Tevatron average and includes 
the run~I reanalysis of the D\O\ lepton plus jets channel\cite{Canelli03}, as
well as first results\cite{Thomson03} from run~II. All correlations and 
a common 0.6~GeV uncertainty due to the conversion from the pole to 
the $\overline{\rm MS}$ mass definition are taken into account.}
\label{mwmt}
\end{figure}

\begin{figure}[t]
\epsfxsize=24pc 
\rotatebox{270.0}{\epsfbox{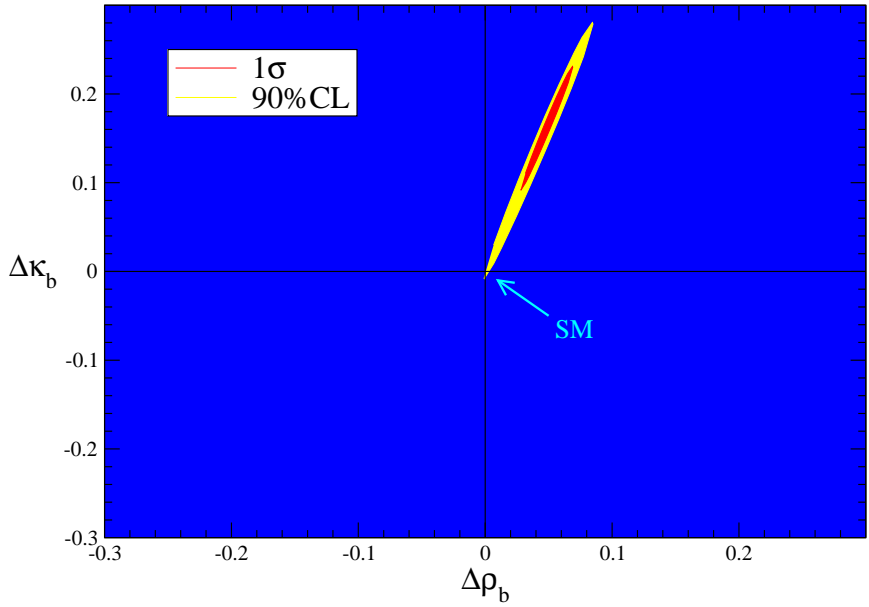}}
\caption{Constraints on new physics contributions to 
$\kappa_b = 1 + \Delta\kappa_b$ (the radiative correction multiplying the weak 
mixing angle entering the $Zb\bar{b}$ vertex) and $\rho_b = 1 + \Delta\rho_b$ 
(the overall normalization of the partial $Z\rightarrow b\bar{b}$ decay width).
$\Delta\kappa_b = \Delta\rho_b = 0$ in the SM by definition.}
\label{Zbb}
\end{figure}

\begin{figure}[t]
\epsfxsize=24pc 
\rotatebox{270.0}{\epsfbox{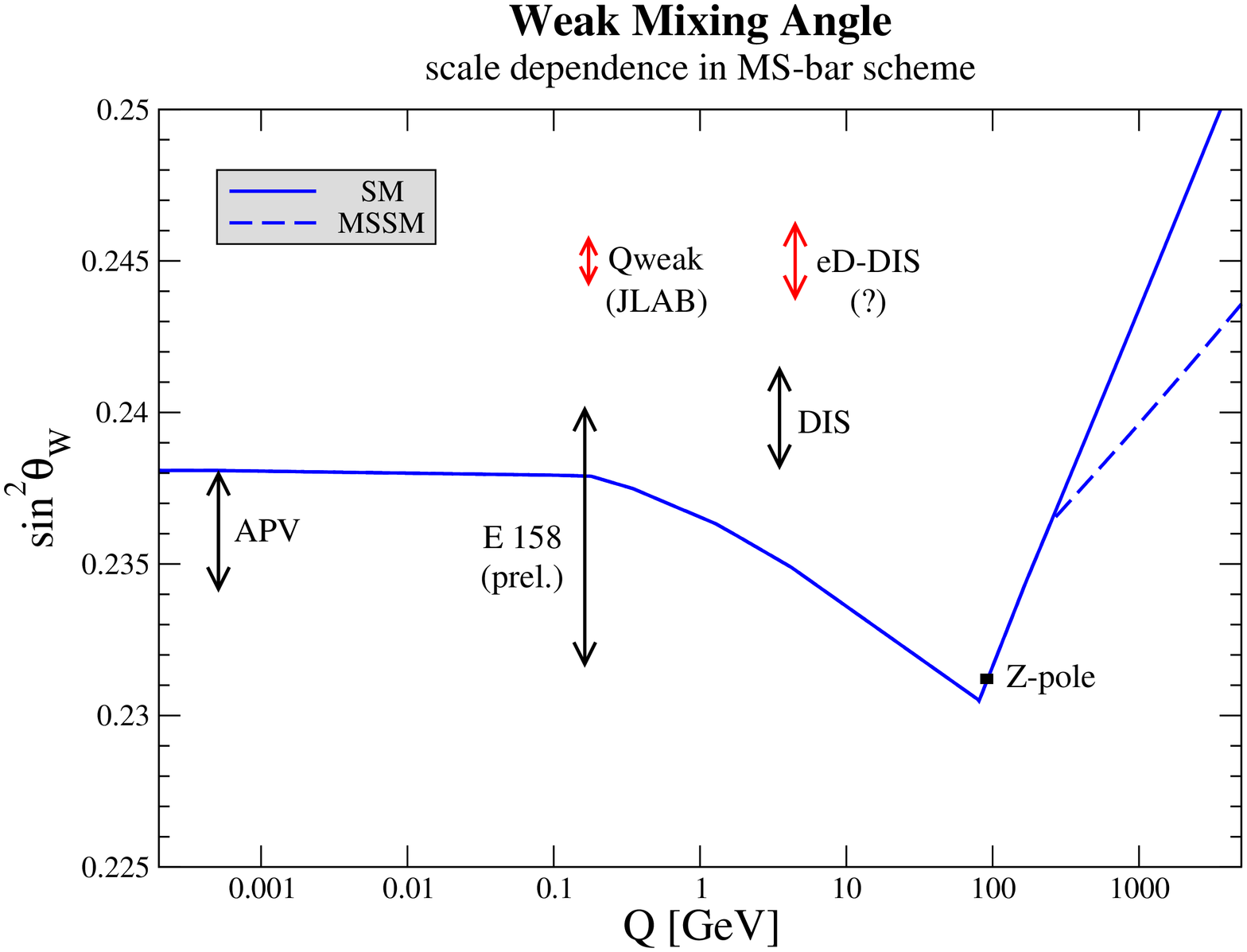}}
\caption[]{Calculated running of the weak mixing angle in the SM, defined in
the $\overline{\rm MS}$ renormalization scheme (the dashed line indicates
the reduced slope typical for the Minimal Supersymmetric Standard Model, MSSM).
Shown are the results from atomic parity violation (Cs\cite{Wood:zq} and 
Tl\cite{Edwards95}), deep inelastic neutrino-nucleon 
scattering\cite{Zeller:2001hh} ($\nu N$-DIS), the preliminary result of 
the first run of polarized M\o ller scattering\cite{Kolomensky03} at SLAC 
(E~158), and from the $Z$-pole\cite{LEPEWWG02}. Qweak is the future measurement
of the weak charge of the proton in low-energy polarized electron-proton 
scattering at JLAB, while eD-DIS refers to a possible polarized 
electron-deuteron experiment (the latter two have arbitrarily chosen vertical 
locations).}
\label{s2w}
\end{figure}

The most fundamental observable related to the weak interaction is the muon
lifetime, $\tau_\mu$.  With the electromagnetic two-loop contribution 
known\cite{vanRitbergen:1998yd}, $\tau_\mu$ can be used unambiguously 
to extract the Fermi constant, $G_F=1.16637(1)\times 10^{-5}\mbox{ GeV}^{-2}$,
where the uncertainty is completely dominated by experiment.  Adding the fine 
structure constant, $\alpha$, one can obtain two relations between 
the intermediate gauge boson masses, $M_{W,Z}$, and the weak mixing 
angle\cite{Degrassi:1990tu},
\be 
\sin^2\hat\theta_W \equiv \shat = \frac{A^2}{M_W^2 (1 - \rwhat)}, \hspace{30pt}
                     \shat (1 - \shat) = \frac{A^2}{M_Z^2 (1 - \rzhat)}.
\label{rzwhat}
\ee
Here $\rwhat$ and $\rzhat$ are electroweak radiative correction parameters (the
caret indicates the $\overline{\rm MS}$ scheme) and the dimensionful quantity
$A^2 = \frac{\pi\alpha}{\sqrt{2} G_F} = (37.2805 \pm 0.0003 \mbox{ GeV})^2$
is known precisely. Most of the $Z$-pole asymmetries are basically measurements
of $\sin^2\theta^{\rm eff}_e = \hat\kappa_e\shat$, where $\hat\kappa_f$ denotes
a flavor dependent form factor. Since furthermore $M_Z$ is known to great 
accuracy, the second Eq.~(\ref{rzwhat}) implies that the $Z$-pole asymmetries 
effectively determine,
\be
    \rzhat = {\alpha\over\pi} \dgama + F_1(m_t^2,M_H,\dots).
\label{rzhat}
\ee
Asymptotically for large top quark masses, $m_t$, the function, $F_1$, grows 
like $m_t^2$. This effect has been absorbed into $G_F$, but now reappears
in $\rzhat$ when $M_Z$ is computed in terms of it. The first Eq.~(\ref{rzwhat})
shows that a determination of the $W$ boson mass can then be used to measure
\be
    \rwhat = {\alpha\over\pi} \dgama + F_2(\ln m_t,M_H,\dots),
\label{rwhat}
\ee
where indeed $F_2$ has a milder $m_t$ dependence.  $F_1$ and $F_2$ are 
complicated functions of the Higgs boson mass, $M_H$, which are asymptotically 
logarithmic.  Eqs.~(\ref{rzhat}) and (\ref{rwhat}) also show that $M_H$ can be
extracted from the precision data only when 
$\dgama/\pi = \alpha^{-1} - \hat\alpha (M_Z)^{-1}$ is known accurately.
Breakdown of the operator product expansion for light quarks, however, 
introduces an uncertainty in $\hat\alpha (M_Z)$ (cf.\ Fig.~\ref{mwmt}). It is 
correlated with the uncertainty in the hadronic two-loop contribution to 
the muon anomalous magnetic moment, $g_\mu - 2$, which is the limiting factor 
in the interpretation of the present world average (dominated by the 1999 and 
2000 data runs\cite{Brown:2001mg} of the E~821 Collaboration at BNL),
$(g_\mu - 2)/2 = (1165920.37 \pm 0.78) \times 10^{-9}$. An evaluation of the SM
prediction\cite{Davier:2003pw} using $e^+e^- \rightarrow$ hadrons cross-section
information (dominated by the recently reanalyzed CMD~2 
data\cite{Akhmetshin:2003zn}) suggests a 1.9~$\sigma$ discrepancy with 
experiment. On the other hand, an alternative analysis\cite{Davier:2003pw} 
based on $\tau$ decay data and isospin symmetry (CVC) indicates no conflict 
(0.7~$\sigma$). Thus, there is also a discrepancy (2.8~$\sigma$) between 
the $2\pi$ spectral functions obtained from the two methods. It is important 
to understand the origin of this difference and to obtain additional 
experimental information. Fortunately, due to the suppression at large $s$ 
(from where the conflict originates) the difference is only 1.7~$\sigma$ as far
as $g_\mu - 2$ is concerned. Note also that part of this difference is due to
older $e^+e^-$ data\cite{Davier:2003pw}. Isospin violating corrections have
been estimated and found to be under control\cite{Cirigliano:2002pv}, where 
the largest effect is due to higher-order electroweak 
corrections\cite{Marciano:vm} but introduces a negligible 
uncertainty\cite{Erler:2002mv}. An additional uncertainty is induced by 
the hadronic three-loop light-by-light type contribution. Averaging the results
from the $e^+e^-$ and $\tau$ based analyzes yields the SM prediction, 
$(g_\mu - 2)/2 = (1165918.83 \pm 0.49) \times 10^{-9}$, where the error 
excludes parametric ones (which are accounted for in the fits). The small 
overall 1.6~$\sigma$ discrepancy between theory and experiment could be due to
fluctuations or underestimates of the theoretical uncertainties. On the other
hand, $g_\mu - 2$ is affected by many types of new physics and the deviation 
might also arise from physics beyond the SM.

\begin{figure}[t]
\epsfxsize=24pc 
\rotatebox{270.0}{\epsfbox{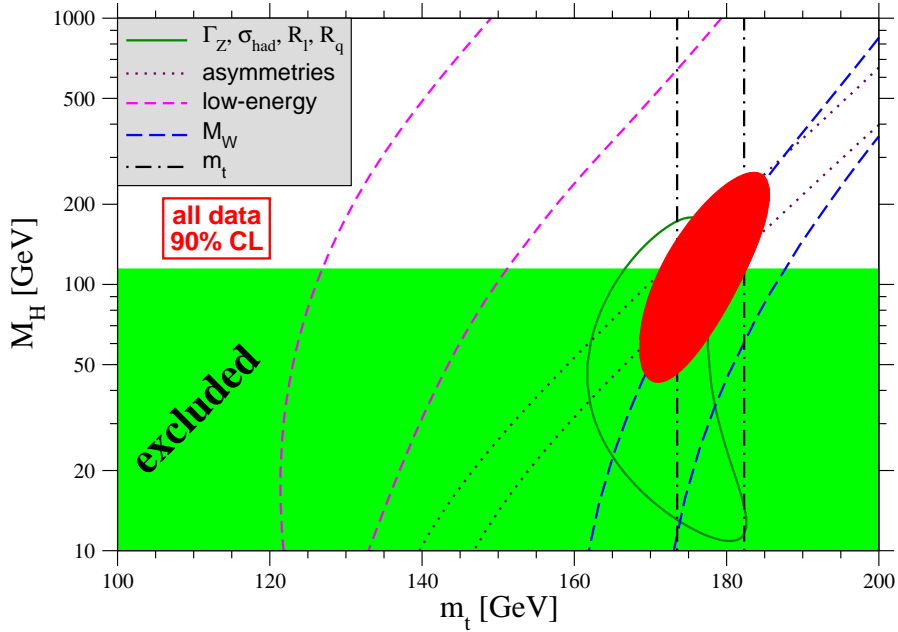}}
\caption[]{One-standard-deviation (39.35\%) uncertainties in $M_H$ as 
a function of $m_t$ for various inputs, and the 90\% CL region 
($\Delta \chi^2 = 4.605$) allowed by all data. $\alpha_s(M_Z) = 0.120$ is
assumed except for the fits including the $Z$-lineshape data. The 95\% direct 
lower limit\cite{LEPHIGGSWG:2003sz} from LEP~2 is also shown.}
\label{mhmt}
\end{figure}

\begin{figure}[t]
\epsfxsize=24pc 
\rotatebox{270.0}{\epsfbox{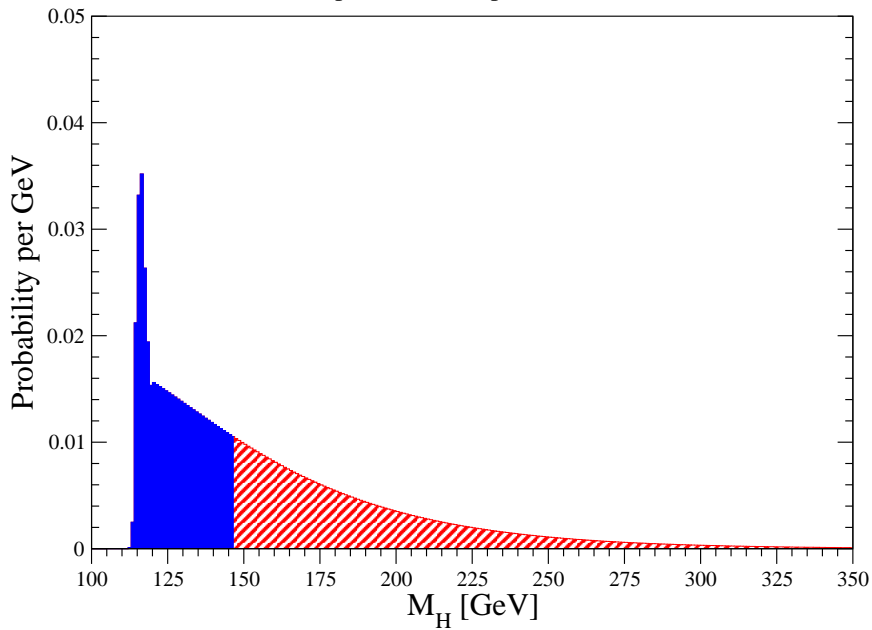}}
\caption[]{Probability density\cite{Erler:2000cr} for $M_H$ obtained by 
combining precision data with the finalized direct search 
results\cite{LEPHIGGSWG:2003sz} at LEP. The peak is due to the candidate Higgs
events seen at LEP~2. The two differently colored and patterned areas contain
50\% probability each.}
\label{mh}
\end{figure}

\begin{figure}[t]
\epsfxsize=24pc 
\rotatebox{270.0}{\epsfbox{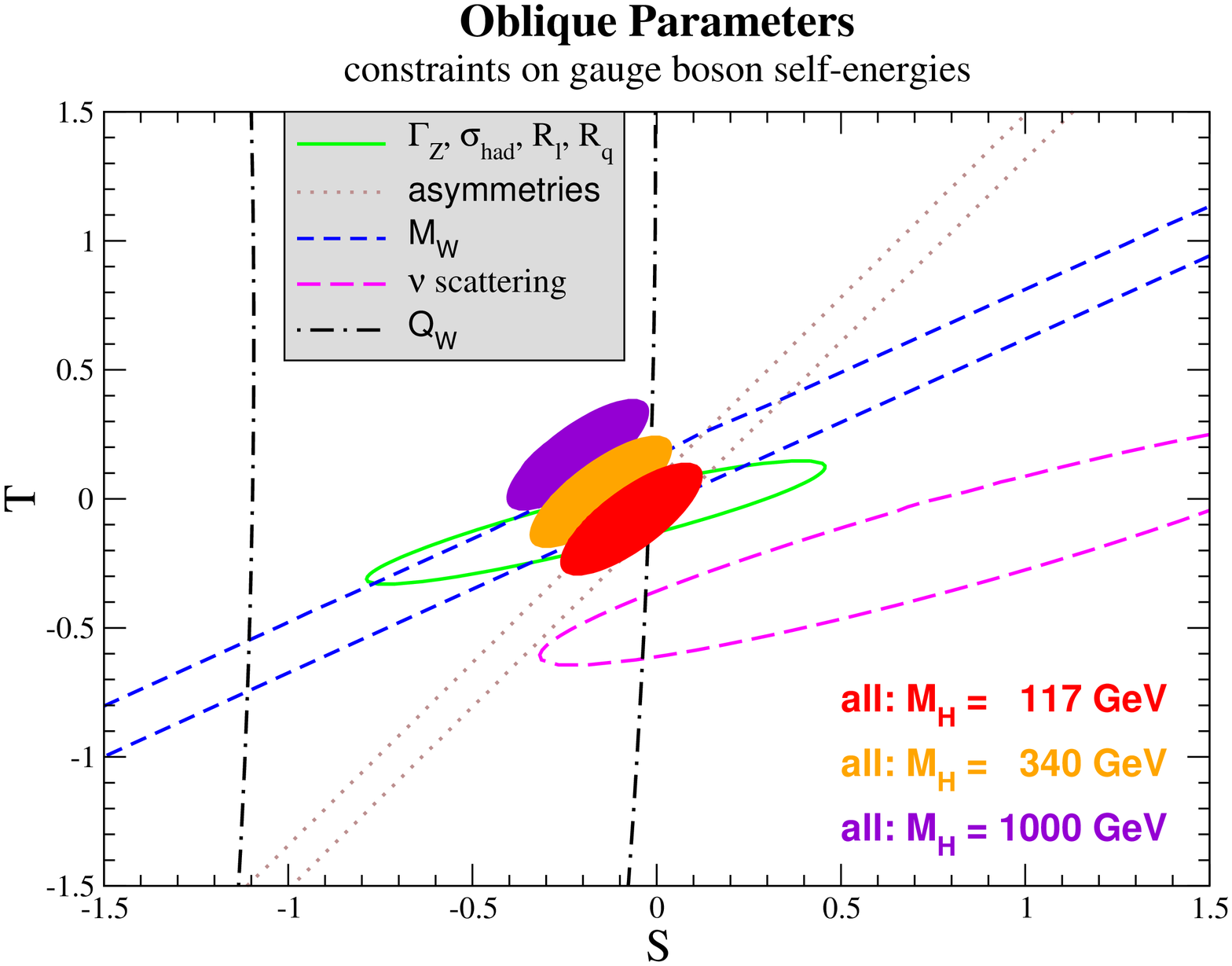}}
\caption{1~$\sigma$ constraints (39.35\%) on $S$ and $T$ from various inputs. 
The contours assume $M_H = 117$~GeV except for the central and upper 90\% CL 
contours allowed by all data, which are for $M_H = 340$~GeV and $1000$~GeV, 
respectively. In all cases $U= 0$ is assumed and $\alpha_s$ is constrained 
using the $\tau$ lifetime as additional input.}
\label{ST}
\end{figure}

Another longstanding deviation is observed\cite{LEPEWWG02} in $Z$ decays to 
$b\bar{b}$ pairs. The forward-backward cross-section asymmetry at LEP~1, 
$A_{FB}^{(b)}$, is 2.2~$\sigma$ below the SM expectation, while the combined 
left-right forward-backward asymmetry, $A_b$, at the SLC and 
the $Z \rightarrow b\bar{b}$ partial width, $R_b$, are in reasonable agreement.
Thus, it is difficult to explain this deviation by new physics effects. As can 
be seen from Fig.~\ref{Zbb}, the model independent form factor determinations 
are marginally consistent with the SM, while large effects (generally too large
to arise from radiative corrections) are needed to explain the central values.
Note, however, that the average of $A_{FB}^{(b)}$ measurements at LEP~2 is also
low (1.6~$\sigma$) and $R_b$ is 2.1~$\sigma$ high. 

The total hadronic cross-section, $\sigma_{\rm had}$, at LEP~2 shows another 
1.7~$\sigma$ excess, which is only marginally significant, but in contrast to 
most other measurements at LEP~2 it is an ${\cal O}(1\%)$ measurement and 
therefore precise enough to be sensitive to TeV scale physics. Interestingly, 
$\sigma_{\rm had}^0$ on top of the $Z$ pole is also 1.9~$\sigma$ high. 
The left-right cross-section asymmetry from the SLD Collaboration for
hadronic\cite{Abe:2000dq} and leptonic\cite{Abe:2000hk} final states show 
a combined deviation of 1.9~$\sigma$ from the SM prediction. In contrast to 
$A_{FB}(b)$ it favors small values of $M_H$, which are excluded by the direct
searches\cite{LEPHIGGSWG:2003sz} at LEP~2, $M_H \geq 114.4$~GeV (95\% CL).
The largest deviation\cite{Zeller:2001hh} (2.9~$\sigma$) is currently in 
the left-handed effective four-Fermi $\nu$-quark coupling, 
$g_L^2 = 1/2 - \shat + 5 \hat{s}^4/9$, while $g_R^2$ agrees with the SM 
prediction. Presently, $g_L^2$ is the most precise measurement of $\shat$ off 
the $Z$-pole (see Fig.~\ref{s2w}).

The various deviations described above notwithstanding, it must be stressed 
that the overall agreement between the data and the SM is excellent. 
The $\chi^2$ per degree of freedom of the global best fit to all data is 
45.5/45, where the probability for a larger $\chi^2$ is 45\%. The data favors 
the range, $M_H = 113^{+56}_{-40}$~GeV, where the central value is very close 
to the lower LEP~2 exclusion limit\cite{LEPHIGGSWG:2003sz} (see 
Fig.~\ref{mhmt}). If one includes the Higgs search 
information\cite{LEPHIGGSWG:2003sz} from LEP, one obtains the probability 
density in Fig.~\ref{mh}. 

Allowing new physics effects in the gauge boson self-energies gives rise (in 
leading order in the new physics) to three parameters\cite{Peskin:1991sw}, $S$,
$T$, and $U$, which are defined to vanish in the SM. Assuming $M_H = 117$~GeV, 
$$ S = - 0.13(10)\;\; [-0.08],\;\;\;\;\;
   T = - 0.17(12)\;\; [+0.09],\;\;\;\;\;
   U =   0.22(13)\;\; [+0.01],$$
where in brackets the shifts are shown for $M_H = 300$~GeV. All deviate by
more than 1~$\sigma$ from zero but this is a correlated effect (see
Fig.~\ref{ST} for $U = 0$).

\section*{Acknowledgments}
It is a pleasure to thank Paul Langacker for collaboration. This work was 
supported by CONACYT (M\'exico) contract 42026--F and by DGAPA--UNAM contract
PAPIIT IN112902.


\begin{thebibliography}{99}
\bibitem{vanRitbergen:1998yd}
T.~van~Ritbergen and R.G.~Stuart, \Journal{\PRL}{82}{488}{1999}.

\bibitem{Degrassi:1990tu}
G.~Degrassi, S.~Fanchiotti, and A.~Sirlin, \Journal{\NPB}{351}{49}{1991}.

\bibitem{Canelli03}
F.~Canelli (D\O), talk presented at CIPANP 2003, New York, NY.

\bibitem{Thomson03}
E.~Thomson (CDF), talk presented at SSI 2003, Menlo Park, CA.

\bibitem{Brown:2001mg}
Muon~g-2~Collaboration: H.N.~Brown {\it et al.}, 
\Journal{\PRL}{86}{2227}{2001}; G.W.~Bennett {\it et al.}, 
\Journal{\PRL}{89}{101804}{2002}.

\bibitem{Davier:2003pw}
M.~Davier, S.~Eidelman, A.~H\"ocker, and Z.~Zhang, {\tt hep-ph/0308213}.

\bibitem{Akhmetshin:2003zn}
CMD~2 Collaboration: R.~R.~Akhmetshin {\it et al.}, {\tt hep-ex/0308008}.

\bibitem{Cirigliano:2002pv}
V.~Cirigliano, G.~Ecker, and H.~Neufeld, {\em JHEP} {\bf 0208}, 002 (2002).

\bibitem{Marciano:vm}
W.J.~Marciano and A.~Sirlin, \Journal{\PRL}{61}{1815}{1988}.

\bibitem{Erler:2002mv}
J.~Erler, {\tt hep-ph/0211345}.

\bibitem{LEPEWWG02}
ALEPH, DELPHI, L3, and OPAL Collaborations, LEP Electroweak Working Group, and
SLD Heavy Flavor Group, {\tt hep-ex/0212036}; \\
M.~Elsing (DELPHI), talk presented at EPS 2003, Aachen, Germany.

\bibitem{Abe:2000dq}
SLD Collaboration: K.~Abe, \etal, \Journal{\PRL}{84}{5945}{2000}.

\bibitem{Abe:2000hk}
SLD Collaboration: K.~Abe, \etal, \Journal{\PRL}{86}{1162}{2001}.

\bibitem{LEPHIGGSWG:2003sz}
ALEPH, DELPHI, L3, and OPAL Collaborations, and LEP Working Group for Higgs 
Boson Searches, \Journal{\PLB}{565}{61}{2003}.

\bibitem{Zeller:2001hh}
NuTeV: G.P.~Zeller \etal, \Journal{\PRL}{88}{091802}{2002}.

\bibitem{Wood:zq}
Boulder: C.S.~Wood \etal, {\em Science} {\bf 275}, 1759 (1997). 

\bibitem{Edwards95}
Oxford: N.H.~Edwards \etal, \Journal{\PRL}{74}{2654}{1995}; \\
Seattle: P.A.~Vetter \etal, \Journal{\PRL}{74}{2658}{1995}.

\bibitem{Kolomensky03} 
Yu.~Kolomensky (E~158), talk presented at DPF 2003, Philadelphia, PA.

\bibitem{Erler:2000cr}
Updated from J.~Erler, \Journal{\PRD}{63}{071301}{2001}.

\bibitem{Peskin:1991sw}
M.E.~Peskin and T.~Takeuchi, \Journal{\PRD}{46}{381}{1992}.

\end{thebibliography}
\end{document}